\documentclass[format=sigconf,review=false,anonymous=false]{acmart}

\renewcommand\footnotetextcopyrightpermission[1]{} 
\usepackage{booktabs}
\usepackage{multirow}
\usepackage[ruled]{algorithm2e}

\usepackage{times}
\usepackage{numprint}
\npdecimalsign{.} 

\usepackage[utf8]{inputenc}
\usepackage[american]{babel}

\usepackage{footnote,xspace,cleveref,graphicx,url,subcaption}

\newcommand{\dmir}{Data Mining and Information Retrieval (DMIR) Group}

\newcommand{\furl}[1]{\footnote{\url{#1}}}

\newcommand{\itrnn}{IT-RNN\xspace}
\newcommand{\dtrnn}{DT-RNN\xspace}

\newcommand{\recsys}{RecSys Challenge 2015\xspace}

\begin{document}
\title{Improving Session Recommendation with Recurrent Neural Networks by Exploiting Dwell Time}

\author{Alexander Dallmann}
\email{dallmann@informatik.uni-wuerzburg.de}
\author{Alexander Grimm}
\email{alexander.grimm@stud-mail.uni-wuerzburg.de}
\author{Christian Pölitz}
\email{poelitz@informatik.uni-wuerzburg.de}
\author{Daniel Zoller}
\email{zoller@informatik.uni-wuerzburg.de}
\affiliation{
  \institution{University of Würzburg}
  \department{\dmir}
  \streetaddress{Am Hubland, Informatik}
  \postcode{97074}
  \country{Germany}
  \city{Würzburg}
}
\author{Andreas Hotho}
\email{hotho@informatik.uni-wuerzburg.de}
\affiliation{
  \institution{University of Würzburg}
  \department{\dmir}
  \streetaddress{Am Hubland, Informatik}
  \postcode{97074}
  \country{Germany}
  \city{Würzburg}
}
\additionalaffiliation{
  \institution{L3S Research Center}
}


\begin{abstract}
Recently, Recurrent Neural Networks (RNNs) have been applied to the task of session-based recommendation.
These approaches use RNNs to predict the next item in a user session based on the previously visited items.
While some approaches consider additional item properties, we argue that item dwell time can be used as an implicit measure of user interest to improve session-based item recommendations.
We propose an extension to existing RNN approaches that captures user dwell time in addition to the visited items and show that recommendation performance can be improved.
Additionally, we investigate the usefulness of a single validation split for model selection in the case of minor improvements and find that in our case the best model is not selected and a fold-like study with different validation sets is necessary to ensure the selection of the best model.
\end{abstract}

\keywords{session recommendation, neural networks, user interest modeling, dwell time, recurrent neural networks}

\maketitle


\section{Introduction}

Today, sales in the Internet are increasing rapidly, but supporting customers while they're shopping is still a challenge as users are  generally hidden by the anonymity of the web. 
One working solution are recommender systems which identify a user's needs by analyzing the shopping history and the user's behavior. 
Therefore, tracking the user becomes an essential tool as it allows to understand the behavior and goals of the customer and ensures the growth of the business. 

Recommendation can be done using explicit information from a user's purchase history, but is then limited to the general past preferences, while the current interest, for example in a new product, is hidden.
However, the current interest of a user can be gleaned from the session and traces left by the user while searching for new products in a web shop. 
These traces left by the user are only implicit feedback which tends to be noisy and difficult to analyze. 
Recently, advanced recommender systems utilize deep learning approaches when analyzing such click traces to predict the way through the shop toward the next purchase.
Typically, a user checks prices and other properties of the next product of interest and investigates similar products.
Therefore, it is a plausible assumption that the next product to buy will be visited by the user while browsing in the shop and can therefore be identified by such learning methods. 

Recent approaches utilize the order of visited items as good indicators for predicting the next click on a page or item. 
Additionally, such a session contains information about the type of products and the frequency of visits. 
When analyzing a user's behavior, one could observe that some products are investigated in more detail than others. 
This different interest is to some extent reflected by the time a user spends with the product before investigating the next one by following a link and is not only expressed by the frequency of the visits within a session.
In this work, we will make use of this additional time information called dwell time \cite{YiBeyondClicks} and show that deep learning networks can make use of this additional information to improve the recommendation quality. 

We propose an extension to a state-of-the-art session-based recurrent neural network model that integrates the item \emph{dwell time} into the model. 
Additionally, we show that care must be taken when conducting a hyper-parameter study to ensure the selection of the best parameters and present a fold-like scheme for selecting the model. 
Finally, results are shown verifying that the dwell time positively impacts recommendation performance. 

The rest of the paper is structured as follows.
In \Cref{sec:related} we discuss related work. 
Next we describe the dataset and give details about the applied preprocessing in \Cref{sec:dataset}.   
This is followed by a description of the studied models in \Cref{sec:model}.
Results of our experiments are shown in \Cref{sec:experiments} followed by a discussion in \Cref{sec:discussion}.
Finally we give a conclusion in \Cref{sec:conclusion}.


\section{Related Work}
\label{sec:related}

Recommendation systems help by suggesting resources (items) in (web) applications based on user and resource preferences~\cite{Bobadilla:2013:RSS:2483330.2483573}. 
Session recommendation systems learn a model of a user's behavior by using a session of events (requested resources, products, pages or more generally items) that are generated by the user.
For example, Item-based recommendations, as in \cite{1167344}, use similar item profiles to recommend new items in the current session.
Session recommendation can also be treated as a sequence learning problem that can be modeled by a Markov Decision Process which predicts possible next events (items) in a given session \cite{Shani:2005:MRS:1046920.1088715} and hence recommend corresponding resources.

These approaches are in contrast to collaborative filtering \cite{Breese:1998:EAP:2074094.2074100} approaches that factorize user information to recommend based on similar user profiles or similarity-based approaches that cluster users~\cite{Herlocker:1999:AFP:312624.312682} or items~\cite{Sarwar:2001:ICF:371920.372071} based on their profiles to extract preferences.

Recently, (deep) neural networks have been used for recommendation systems \cite{Salakhutdinov:2007:RBM:1273496.1273596}. 
These networks learn feature representations of items, users or whole sessions. 
\cite{Wang:2015:CDL:2783258.2783273}, for example, learn latent feature representations of content information with stacked denoising autoencoders, \cite{Oord:2013:DCM:2999792.2999907} apply convolutional neural networks on audio content for music recommendation, \cite{Greenstein-Messica:2017:SRU:3025171.3025197} use item embeddings and \cite{Covington:2016:DNN:2959100.2959190} use embeddings of videos, search query terms and user features on Youtube. 
Different combinations of network architectures have been proposed in the literature. For instance, \cite{DBLP:journals/corr/ChengKHSCAACCIA16} investigate combining deep (neural network) or wide (linear models) architectures to capture generalization capabilities of a deep neural network together with modeling strength of sparse feature interaction of linear models.
\cite{Twardowski:2016:MCI:2959100.2959162} use a feed forward neural network to encode item information and an RNN to encode session information for session-based recommendations.
\cite{Song:2016:MDL:2911451.2914726} explicitly model temporal behavior by RNNs together with user and item features using feed forward nets to perform temporal recommendations. 

The sequential nature of most of the systems for session-based recommendations makes recurrent networks a good model candidate. 
Recurrent Neural Networks (RNNs) can be used to make predictions on sequential input data.
Recently, \citeauthor{Hidasigru4rec} proposed a sequence-to-sequence RNN model for predicting the user session from the sequence of clicks~\cite{Hidasigru4rec}.
In contrast, \citeauthor{TanSessRec} use an RNN model that predicted only the last click in a session and studied the impact of item embeddings and data augmentation on the prediction performance~\cite{TanSessRec}. 
Considering a large set of additional features to the items, \cite{Hidasi:2016:PRN:2959100.2959167} propose parallel RNNs.
Further, temporal features like the date or time of the current session have been used in \cite{Dias:2013:IMR:2571275.2572910} to recommend music. 
In contrast to this approach, we use the dwell time on a per item basis as a dwell time profile for the session. 
Dwell time as a user feature for personalized recommendations is investigated in \cite{YiBeyondClicks}. 
The dwell time is used to express preferences for a certain time. 
Hence, the longer the dwell time, the more relevant the item is to a user. 
Here, the dwell times are only used to rank the items for recommendations.

\section{Dataset and Preprocessing}
\label{sec:dataset}
We evaluate all models on the open \recsys dataset.
The dataset contains click events from an e-commerce store that can be aggregated to user sessions and was published as part of the RecSys Challenge 2015\footnote{\url{https://recsys.acm.org/recsys15/challenge/}}.
Overall, the dataset contains \numprint{9249729} sessions with \numprint{33003944} clicks and \numprint{52739} unique items.
\Cref{tab:yoochoose} shows the different attributes that are collected for each click.

\begin{table}[htp]
	\caption{Relevant properties of the RecSys15 dataset.
	}
	\label{tab:yoochoose}
	\begin{tabular}{c l}
		\toprule
		Property & Description \\
		\midrule
		sid & an id for the session this click event belongs to\\
		timestamp & the time the click event was recorded\\
		item & the id for the item that was clicked\\
		category & the item category\\
		\bottomrule
	\end{tabular}
\end{table}

In general, we apply the same preprocessing steps as described in \cite{Hidasigru4rec}.
More specifically, the dataset is split into a train and test set and the test set contains all sessions of the last day. During the analysis of the evaluation in the parameter study, with split the dataset as follows.
For a fold-like validation scheme, we create six splits of the training set with each split using one of the last six days as the validation set.
Additionally, all sequences with length $l<2$ and items with a support $sup<5$ are removed. We also remove items from the validation and test sets that don't occur in the respective training sets.
As proposed in \cite{TanSessRec}, the session length ($l$) is restricted and chosen so that $max\left(l\right)=16$ and all sessions for which $l > 16$ are removed. This captures approximately $98\%$ of all sessions.
The properties of the train and test sets after preprocessing are listed in \Cref{tab:yoochoosePre}.  
We also produce a second dataset with augmented sessions as described in \cite{TanSessRec}, where for each session every prefix is added as a separate example.
 
\begin{table}[htp]
	\caption{Number of session, items and the avg. session length of the training and test sets.}
	\label{tab:yoochoosePre}
	\begin{tabular}{c c c}
		\toprule
		  & Train & Test  \\
		\midrule
		\#sessions & 12,864,743 & 30,484 \\
		\#items & 53,308,101 & 136,150 \\		
		$\overline{session\_length}$ & 4.14 & 4.47 \\
		\bottomrule
	\end{tabular}
\end{table}
\section{Methods}
\label{sec:model}
In this section, we first give a description of the model in \Cref{sec:probset}. 

Then we shortly describe the previously proposed approaches for session-based recommendation using RNNs in \Cref{sec:itrnn}. 
Next, we present our extension for integrating dwell time information in \Cref{sec:dtrnn}. 

\subsection{Problem Setting}
\label{sec:probset}
We consider the task of predicting the next item in a session based on its previous items in the same session.
Let a session $S$ be an ordered series of clicks $S=(c_{1}, c_{2}, ..., c_{k})$ with length $k$ and a click $c$ be a tuple $c=(i, t)$, where $i\in I$ is a unique identifier for the clicked item and $t$ contains a timestamp when this click occurred.
Our task is then to fit a Model $M$ defined as $\textbf{y}=M(\textbf{x})$ that computes a probability distribution $\textbf{y} \in \mathbf{R}^{|I|}$ from an input sequence $\textbf{x} = (c_1, c_2, ..., c_{k})$.

\subsection{Recurrent Neural Networks Based on Item Sequences}
\label{sec:itrnn}
Recently, different RNN architectures have been successfully proposed for this task.
In \cite{Hidasigru4rec}, a sequence-to-sequence RNN with Gated Recurrent Unit (GRU) layers is proposed that uses a session-parallel scheme to process the input sessions.
Another model that, in contrast, predicts only the next item $i_{k}$ in a sequence $[i_{1}, i_{2}, \cdots, i_{k-1}]$ is studied in \cite{TanSessRec}. 
In order to facilitate the prediction of every item in a session, a data augmentation scheme is introduced where every session prefix is used as a distinct sample.
Additionally, item embeddings are used as inputs and improvements over \cite{Hidasigru4rec} in terms of Recall@20 and MRR@20 are reported.
 
\begin{figure}[htp]
    \includegraphics[width=0.21\textwidth]{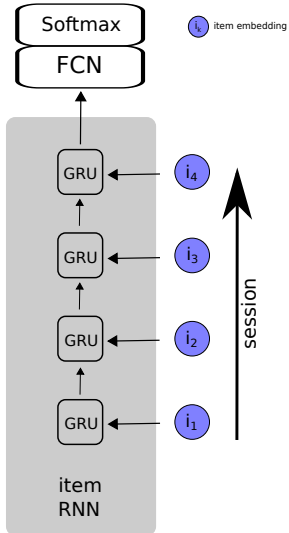}
    \caption{An illustration of the IT-RNN model for predicting the next item in a sequence given only the items as input.}
	\label{fig:itrnn}	
\end{figure}

Because of the better adaptability to our task and  the reported results we use a model similar to the
one described in \cite{TanSessRec} for solely item-based recommendation and call it IT-RNN.
Specifically, we use item embeddings as input to an RNN consisting of GRU layers and predict the next item $i_{k}$ in a fixed sequence of length $l=k-1$.
The model produces a probability distribution over all items using a single fully-connected layer followed by softmax as the final layer. In preliminary experiments, it turns out that this setting is superior to all the other tested configurations.
An illustration of the IT-RNN model its shown in \Cref{fig:itrnn}.

\subsection{Recurrent Neural Network Combining Item and Dwell Time Sequences}
\label{sec:dtrnn}

The described model predicts the next item in the sequences only based on previously visited items. 
However, the amount of time a user spends with an item (dwell time) can be an important metric for user engagement and interest \cite{YiBeyondClicks}.

For every session, we compute the dwell time for item $i_{k}$ in a session as $d_{k} = t_{k+1} - t_{k}$ and get an aligned sequence of dwell times that characterizes the user interest over the items. 

While dwell times are computed from timestamps and are therefore measured in milliseconds, such a high resolution is unlikely to be interesting when measuring user interest.

\begin{figure}[htp]
    \includegraphics[width=0.5\textwidth]{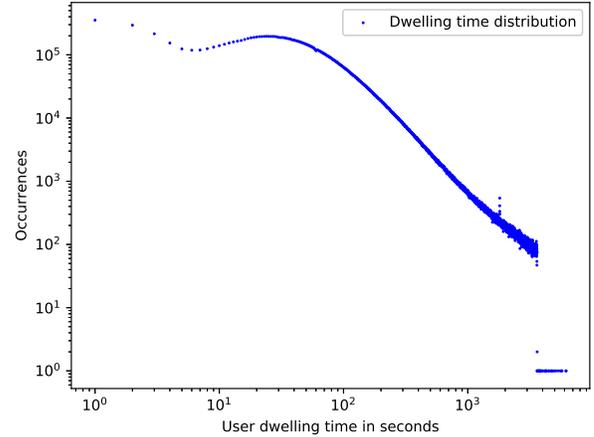}
    \caption{Distribution of dwell times in the preprocessed dataset.}
  \label{fig:dt_dist}
\end{figure}

Therefore, we propose a simple bucketing technique, where dwell times are rounded to the nearest second. 
The resulting distribution of the rounded dwell times for the \recsys dataset is shown in \Cref{fig:dt_dist}.
It already shows some interesting artifacts, like the two peaks.
One with a very short dwell time and one at approximately 35s.
After that, we observe an expected decrease in the log-log plot which only contains some minor exceptions. 
Using each second as a label, the dwell time for each item in a session can be encoded as a discrete class.
 
In most settings the captured dwell time will have some upper limit that is enforced by the consumed service, e.g. a connection timeout.
However, since the upper bound is arbitrary, the representation can become overly sparse and computations could become inefficient. Furthermore, we assume that e.g. the peak with a very short time somehow encodes skipping the visited pages. The encoding of such an information is important, but rather difficult. 
In order to encode the hidden information of the distribution and to avoid problems with dwell times of arbitrary length, we encode the dwell time classes into continuous lower dimensional embeddings.

\begin{figure}[htp]
    \includegraphics[width=0.45\textwidth]{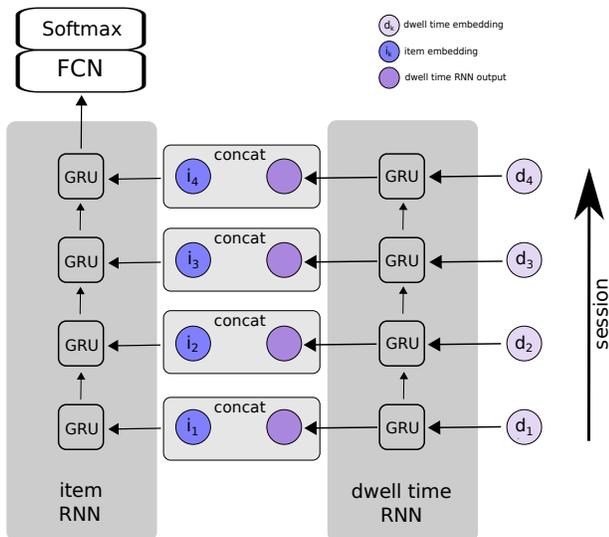}
    \caption{The DT-RNN network architecture.}
	\label{fig:dtrnn}	
\end{figure}

To capture a sense of user interest over a session, we use the sequence of dwell time embeddings as input to an RNN with a GRU layer.
The output of the RNN at each step in the sequence is concatenated with the embedded item at the same step and the result is used as input to the \itrnn part (Please keep in mind that the concatenated vectors are of different size).
This way the interest a user expresses over a session can act as a weight that boosts or dampens the influence an item has on the recommendation of the next item. We call this model DT-RNN, an illustration is shown in \Cref{fig:dtrnn}.

\section{Experiments}
\label{sec:experiments}
In this section, the experiments and results for our model are presented. 
All experiments are conducted using the dataset described in \Cref{sec:dataset}.
The models are implemented in Tensorflow\footnote{\url{https://www.tensorflow.org}} and are trained for six epochs over the training data on a NVidia GTX 1080 Founders Edition graphics card. 
Batch size is set to the maximum possible value for each model depending on the model size. Adam \cite{Kingma2014AdamAM} is used as the optimizer with the default settings provided by Tensorflow.
For evaluation, Recall@20 and MRR@20 (Mean Reciprocal Rank) are calculated on the test set for each model and epoch, and evaluation scores of the best epoch are reported.

We want to study whether our model can use the dwell time information in addition to the item sequence to more accurately predict a user's next click and hence improve recommendation performance and user experience.
Therefore, we first investigate the effects of item embeddings and augmentation on the item sequence based \itrnn and fix the relevant hyperparameters for both models.
Next, we perform a parameter study on the remaining parameters of the \dtrnn model and present the results.

\subsection{Parameter Selection for \itrnn}
Results in \cite{Hidasigru4rec} and \cite{TanSessRec} show that the best performance is achieved with an RNN layer size of $100$ when comparing to a larger network with a layer of size $1000$.
We therefore fix the layer for \itrnn to a similar size of $128$.

Both one-hot codings and embeddings have been studied for encoding the items as input to the RNN with mixed results in \cite{Hidasigru4rec} and \cite{TanSessRec}.
Hence, we experiment with both embeddings of size $128$ and one-hot codings and find that embeddings clearly outperform one-hot codings in our setup. 

{
\nprounddigits{4}
\begin{table}
	\caption{Comparison of one-hot codings versus embeddings for \itrnn with $it\_rnn\_size=128$ and $item\_em\_size=128$. Both networks are first trained without augmenation and then with augmentation and Recall@20 is computed on the augmented test set in both cases.}
	\label{tab:repr}
	\begin{tabular}{l l l}
		\toprule
		  & one-hot & embedding\\
		\midrule
		with augmentation & $\numprint{0.6484}$ & \numprint{0.6871} \\
		without augmentation & $\numprint{0.6294}$ & \numprint{0.6533} \\
		\bottomrule
	\end{tabular}
\end{table}
}

Next, we experiment with augmented sessions as introduced in \cite{TanSessRec} and find that training with augmented datasets improves the performance of the network on the test set. \Cref{tab:repr} shows the Recall@20 values with and without augmentation for both embedded and one-hot coded items.

Based on the results we fix the size of the item embeddings to $item\_em\_size=128$ and use the augmented dataset for training and evaluation.
The evaluation results on the test set for \itrnn are reported in \Cref{tab:it_rnn} and will be used as a baseline to judge whether the proposed \dtrnn model can improve recommendation performance.

{
\nprounddigits{4}
\begin{table}
	\caption{Results for \itrnn with $item\_em\_size = 128$ and $it\_rnn\_size = 128$.}
	\label{tab:it_rnn}
	\begin{tabular}{l l l}
		\toprule
		Method & R@20 & MRR@20 \\
		\midrule
		\itrnn & $\numprint{0.6871}$ & $\numprint{0.2829}$ \\
		\bottomrule
	\end{tabular}
\end{table}
}

\subsection{Influence of Dwell Time}

We want to study the effect of dwell time on the performance compared to the \itrnn model.
Hence, we fix the parameters for the \itrnn part and only conduct the parameter studies for the remaining parameters of the \dtrnn model.
We perform a grid search for the dwell time embedding size $dt\_em\_size$ and the dwell time rnn size $dt\_rnn\_size$ with values for $dt\_em\_size$ chosen from $[4, 8, 16, 32]$ and values for $dt\_rnn\_size$ selected from $[4, 8, 16, 32, 64, 128]$.

First, we use the second to last day (sixth) day as the validation set and the remaining days for training (as given in \Cref{sec:dataset}) and run a grid search with the former specified values.
Based on the evaluation on the validation set, $dt\_rnn\_size=16$ and $dt\_em\_size=8$ are selected as the best performing hyper-parameters and a model using these parameters is trained on the full training set and evaluated on the test set.
The results are listed as \textit{DT-RNN} in \Cref{tab:dtv_rnn}.
A comparison between the results of \itrnn in \Cref{tab:it_rnn} and \dtrnn in \Cref{tab:dtv_rnn} with the selected parameters shows no improvement over \itrnn.

From results in \cite{YiBeyondClicks} we know that we should not expect big improvements. 
However, if the expected improvements are only small, it is possible that using an arbitrary and arguably small validation set might account for the selection of the wrong model, which hides a possible gain. 
We therefore investigate whether a parameter setting in our grid exists that shows significant improvement over the plain \itrnn by conducting a grid search on the full training set using the test set as the validation set. 
Surprisingly, the best performing setting is different from before and is listed in \Cref{tab:dtv_rnn} as \textit{DT-RNN*}.
A Wilcoxon signed rank test showed that the improvement over \itrnn in Rec@20 is statistically significant with $p<0.01$, while
no significant improvement for MRR@20 can be observed.

The different model selection results could be caused by parameter overfitting or by an arbitrary small validation set. 
To decide this question, we conduct a third parameter study and train both parameter settings in \Cref{tab:dtv_rnn} on six different train-validation splits as described in \Cref{sec:dataset}.
\Cref{tab:folds} shows the results for the six different splits.
Only two of them, including our initial validation set, show a better performance with the initially chosen parameter settings ($dt\_rnn\_size=16$), while the majority of four splits shows higher performance for the optimal setting ($dt\_rnn\_size=32$).
Calculating the average Rec@20 values for both models over all splits shows that on average the best performing model \textit{DT-RNN*} should have been selected with $dt\_rnn\_size=32$ which results in a significant improvement of our approach.

{
\nprounddigits{4}
\begin{table}
	\caption{Metrics on the test set calculated for the settings selected by the parameter study $DT-RNN$ and the best settings on the test set $DT-RNN^{*}$.}
	\label{tab:dtv_rnn}
	\begin{tabular}{c c c c c}
		\toprule
		Method & $dt\_em\_size$ & $dt\_rnn\_size$ & Rec@20 & MRR@20 \\
		\midrule
		$DT-RNN$ & 16 & 8 & $\numprint{0.6872702205882353}$ & $\numprint{0.2809558311429833}$ \\
		$DT-RNN^{*}$ & 32 & 8 & $\numprint{0.6926}$ & $\numprint{0.2836}$ \\
		\bottomrule
	\end{tabular}
\end{table}
}

{
\nprounddigits{4}
\begin{table}
	\caption{Rec@20 values with different days chosen as the validation set. The hyper parameters selected by a parameter study $dt\_em\_size=16$ on a fixed validation day (here day $6^{*}$) are compared to those with the best performance on the test set $dt\_em\_size=32$.}
	\label{tab:folds}
	\begin{tabular}{c c c c}
		\toprule
		Day & $dt\_rnn\_size$ & \multicolumn{2}{c}{Rec@20} \\ 
		& & $dt\_em\_size=16$ & $dt\_em\_size=32$ \\
		\midrule
		1 & 8 & \numprint{0.636482} & \textbf{\numprint{0.639205}}\\
		2 & 8 & \numprint{0.629568} & \textbf{\numprint{0.631944}}\\
		3 & 8 & \numprint{0.661093} & \textbf{\numprint{0.665537}}\\
		4 & 8 & \textbf{\numprint{0.673821}} & \numprint{0.673638}\\
		5 & 8 & \numprint{0.684637} & \textbf{\numprint{0.688529}}\\
		6* & 8 & \textbf{\numprint{0.666146}} & \numprint{0.663411}\\
		\midrule
		$\overline{Rec@20}$ & & \numprint{0.65862} & \textbf{\numprint{0.66035}} \\
		\bottomrule
	\end{tabular}
\end{table}
}


\section{Discussion}
\label{sec:discussion}

Our experiments show that the proposed extension can exploit dwell times to boost the performance in terms of Rec@20. 
Unfortunately, the gain is rather limited and seems to be hidden by other effects, as our analysis reveals. 

Selecting the correct model in a parameter study is not straightforward and is also influenced by the selected parts of the dataset used for training and validation. 
Conducting the study on a small and limited validation set can yield the wrong parameter combination if the improvement is small, as in our case.
One solution can be to use several folds or a complete leave-one-out scheme. 
We experimented successfully with using different days as validation sets, but a more detailed analysis is also needed for experiments relying on the same split of the data as we used to check the generality of the results. 

The used dataset split is also of interest in other settings, as a poorly chosen validation set could lead to results which are not applicable in general. 
This could be caused by several reasons. 
When dealing with temporal data, such data typically contains seasonally or weekly effects, i.e. buy patterns which show a significantly different behavior on different time slots. 
In this case, one can no longer assume that the distribution of the data is the same, which could be the reason for the observed results in our experiments. 
Therefore, care must be taken when evaluating models on small and timely restricted datasets.

\section{Conclusion}
\label{sec:conclusion}
In this work, we showed that the performance of an RNN for session-based recommendation can be improved by integrating item dwell times into the model. 
Additionally, we demonstrated that using a single validation set for a parameter study can lead to a sub optimal choice of parameters if the performance gain is only small.
Furthermore, we showed that using a fold-like scheme with several validation sets can help to find the optimal parameters in this case. 

In future work, we would like to provide a more detailed analysis of temporal effects in datasets comprised of user sessions on model selection in parameter studies and find general guidelines for performing these studies.
Additionally, we want to investigate the effectiveness of our model on different datasets from various domains.
Finally, we would like to evaluate the effectiveness of incorporating user dwell time as a measure of user interest in other settings, e.g. buy event prediction.

\bibliographystyle{ACM-Reference-Format}
\bibliography{bibliography}


\begin{thebibliography}{00}


\ifx \showCODEN    \undefined \def \showCODEN     #1{\unskip}     \fi
\ifx \showDOI      \undefined \def \showDOI       #1{{\tt DOI:}\penalty0{#1}\ }
  \fi
\ifx \showISBNx    \undefined \def \showISBNx     #1{\unskip}     \fi
\ifx \showISBNxiii \undefined \def \showISBNxiii  #1{\unskip}     \fi
\ifx \showISSN     \undefined \def \showISSN      #1{\unskip}     \fi
\ifx \showLCCN     \undefined \def \showLCCN      #1{\unskip}     \fi
\ifx \shownote     \undefined \def \shownote      #1{#1}          \fi
\ifx \showarticletitle \undefined \def \showarticletitle #1{#1}   \fi
\ifx \showURL      \undefined \def \showURL       #1{#1}          \fi
\providecommand\bibfield[2]{#2}
\providecommand\bibinfo[2]{#2}
\providecommand\natexlab[1]{#1}
\providecommand\showeprint[2][]{arXiv:#2}

\bibitem[\protect\citeauthoryear{Bobadilla, Ortega, Hernando, and
  Guti{\'e}Rrez}{Bobadilla et~al\mbox{.}}{2013}]%
        {Bobadilla:2013:RSS:2483330.2483573}
\bibfield{author}{\bibinfo{person}{J. Bobadilla}, \bibinfo{person}{F. Ortega},
  \bibinfo{person}{A. Hernando}, {and} \bibinfo{person}{A. Guti{\'e}Rrez}.}
  \bibinfo{year}{2013}\natexlab{}.
\newblock \showarticletitle{Recommender Systems Survey}.
\newblock \bibinfo{journal}{{\em Know.-Based Syst.\/}}  \bibinfo{volume}{46}
  (\bibinfo{date}{July} \bibinfo{year}{2013}), \bibinfo{pages}{109--132}.
\newblock
\showISSN{0950-7051}
\showDOI{%
\url{https://doi.org/10.1016/j.knosys.2013.03.012}}


\bibitem[\protect\citeauthoryear{Breese, Heckerman, and Kadie}{Breese
  et~al\mbox{.}}{1998}]%
        {Breese:1998:EAP:2074094.2074100}
\bibfield{author}{\bibinfo{person}{John~S. Breese}, \bibinfo{person}{David
  Heckerman}, {and} \bibinfo{person}{Carl Kadie}.}
  \bibinfo{year}{1998}\natexlab{}.
\newblock \showarticletitle{Empirical Analysis of Predictive Algorithms for
  Collaborative Filtering}. In \bibinfo{booktitle}{{\em Proceedings of the
  Fourteenth Conference on Uncertainty in Artificial Intelligence}} {\em
  (\bibinfo{series}{UAI'98})}. \bibinfo{publisher}{Morgan Kaufmann Publishers
  Inc.}, \bibinfo{address}{San Francisco, CA, USA}, \bibinfo{pages}{43--52}.
\newblock
\showISBNx{1-55860-555-X}
\showURL{%
\url{http://dl.acm.org/citation.cfm?id=2074094.2074100}}


\bibitem[\protect\citeauthoryear{Cheng, Koc, Harmsen, Shaked, Chandra, Aradhye,
  Anderson, Corrado, Chai, Ispir, Anil, Haque, Hong, Jain, Liu, and Shah}{Cheng
  et~al\mbox{.}}{2016}]%
        {DBLP:journals/corr/ChengKHSCAACCIA16}
\bibfield{author}{\bibinfo{person}{Heng{-}Tze Cheng}, \bibinfo{person}{Levent
  Koc}, \bibinfo{person}{Jeremiah Harmsen}, \bibinfo{person}{Tal Shaked},
  \bibinfo{person}{Tushar Chandra}, \bibinfo{person}{Hrishi Aradhye},
  \bibinfo{person}{Glen Anderson}, \bibinfo{person}{Greg Corrado},
  \bibinfo{person}{Wei Chai}, \bibinfo{person}{Mustafa Ispir},
  \bibinfo{person}{Rohan Anil}, \bibinfo{person}{Zakaria Haque},
  \bibinfo{person}{Lichan Hong}, \bibinfo{person}{Vihan Jain},
  \bibinfo{person}{Xiaobing Liu}, {and} \bibinfo{person}{Hemal Shah}.}
  \bibinfo{year}{2016}\natexlab{}.
\newblock \showarticletitle{Wide {\&} Deep Learning for Recommender Systems}.
\newblock \bibinfo{journal}{{\em CoRR\/}}  \bibinfo{volume}{abs/1606.07792}
  (\bibinfo{year}{2016}).
\newblock
\showURL{%
\url{http://arxiv.org/abs/1606.07792}}


\bibitem[\protect\citeauthoryear{Covington, Adams, and Sargin}{Covington
  et~al\mbox{.}}{2016}]%
        {Covington:2016:DNN:2959100.2959190}
\bibfield{author}{\bibinfo{person}{Paul Covington}, \bibinfo{person}{Jay
  Adams}, {and} \bibinfo{person}{Emre Sargin}.}
  \bibinfo{year}{2016}\natexlab{}.
\newblock \showarticletitle{Deep Neural Networks for YouTube Recommendations}.
  In \bibinfo{booktitle}{{\em Proceedings of the 10th ACM Conference on
  Recommender Systems}} {\em (\bibinfo{series}{RecSys '16})}.
  \bibinfo{publisher}{ACM}, \bibinfo{address}{New York, NY, USA},
  \bibinfo{pages}{191--198}.
\newblock
\showISBNx{978-1-4503-4035-9}
\showDOI{%
\url{https://doi.org/10.1145/2959100.2959190}}


\bibitem[\protect\citeauthoryear{Dias and Fonseca}{Dias and Fonseca}{2013}]%
        {Dias:2013:IMR:2571275.2572910}
\bibfield{author}{\bibinfo{person}{Ricardo Dias} {and}
  \bibinfo{person}{Manuel~J. Fonseca}.} \bibinfo{year}{2013}\natexlab{}.
\newblock \showarticletitle{Improving Music Recommendation in Session-Based
  Collaborative Filtering by Using Temporal Context}. In
  \bibinfo{booktitle}{{\em Proceedings of the 2013 IEEE 25th International
  Conference on Tools with Artificial Intelligence}} {\em
  (\bibinfo{series}{ICTAI '13})}. \bibinfo{publisher}{IEEE Computer Society},
  \bibinfo{address}{Washington, DC, USA}, \bibinfo{pages}{783--788}.
\newblock
\showISBNx{978-1-4799-2971-9}
\showDOI{%
\url{https://doi.org/10.1109/ICTAI.2013.120}}


\bibitem[\protect\citeauthoryear{Greenstein-Messica, Rokach, and
  Friedman}{Greenstein-Messica et~al\mbox{.}}{2017}]%
        {Greenstein-Messica:2017:SRU:3025171.3025197}
\bibfield{author}{\bibinfo{person}{Asnat Greenstein-Messica},
  \bibinfo{person}{Lior Rokach}, {and} \bibinfo{person}{Michael Friedman}.}
  \bibinfo{year}{2017}\natexlab{}.
\newblock \showarticletitle{Session-Based Recommendations Using Item
  Embedding}. In \bibinfo{booktitle}{{\em Proceedings of the 22Nd International
  Conference on Intelligent User Interfaces}} {\em (\bibinfo{series}{IUI
  '17})}. \bibinfo{publisher}{ACM}, \bibinfo{address}{New York, NY, USA},
  \bibinfo{pages}{629--633}.
\newblock
\showISBNx{978-1-4503-4348-0}
\showDOI{%
\url{https://doi.org/10.1145/3025171.3025197}}


\bibitem[\protect\citeauthoryear{Herlocker, Konstan, Borchers, and
  Riedl}{Herlocker et~al\mbox{.}}{1999}]%
        {Herlocker:1999:AFP:312624.312682}
\bibfield{author}{\bibinfo{person}{Jonathan~L. Herlocker},
  \bibinfo{person}{Joseph~A. Konstan}, \bibinfo{person}{Al Borchers}, {and}
  \bibinfo{person}{John Riedl}.} \bibinfo{year}{1999}\natexlab{}.
\newblock \showarticletitle{An Algorithmic Framework for Performing
  Collaborative Filtering}. In \bibinfo{booktitle}{{\em Proceedings of the 22Nd
  Annual International ACM SIGIR Conference on Research and Development in
  Information Retrieval}} {\em (\bibinfo{series}{SIGIR '99})}.
  \bibinfo{publisher}{ACM}, \bibinfo{address}{New York, NY, USA},
  \bibinfo{pages}{230--237}.
\newblock
\showISBNx{1-58113-096-1}
\showDOI{%
\url{https://doi.org/10.1145/312624.312682}}


\bibitem[\protect\citeauthoryear{Hidasi, Karatzoglou, Baltrunas, and
  Tikk}{Hidasi et~al\mbox{.}}{2015}]%
        {Hidasigru4rec}
\bibfield{author}{\bibinfo{person}{Bal{\'a}zs Hidasi},
  \bibinfo{person}{Alexandros Karatzoglou}, \bibinfo{person}{Linas Baltrunas},
  {and} \bibinfo{person}{Domonkos Tikk}.} \bibinfo{year}{2015}\natexlab{}.
\newblock \showarticletitle{Session-based Recommendations with Recurrent Neural
  Networks}.
\newblock \bibinfo{journal}{{\em CoRR\/}}  \bibinfo{volume}{abs/1511.06939}
  (\bibinfo{year}{2015}).
\newblock


\bibitem[\protect\citeauthoryear{Hidasi, Quadrana, Karatzoglou, and
  Tikk}{Hidasi et~al\mbox{.}}{2016}]%
        {Hidasi:2016:PRN:2959100.2959167}
\bibfield{author}{\bibinfo{person}{Bal\'{a}zs Hidasi}, \bibinfo{person}{Massimo
  Quadrana}, \bibinfo{person}{Alexandros Karatzoglou}, {and}
  \bibinfo{person}{Domonkos Tikk}.} \bibinfo{year}{2016}\natexlab{}.
\newblock \showarticletitle{Parallel Recurrent Neural Network Architectures for
  Feature-rich Session-based Recommendations}. In \bibinfo{booktitle}{{\em
  Proceedings of the 10th ACM Conference on Recommender Systems}} {\em
  (\bibinfo{series}{RecSys '16})}. \bibinfo{publisher}{ACM},
  \bibinfo{address}{New York, NY, USA}, \bibinfo{pages}{241--248}.
\newblock
\showISBNx{978-1-4503-4035-9}
\showDOI{%
\url{https://doi.org/10.1145/2959100.2959167}}


\bibitem[\protect\citeauthoryear{Kingma and Ba}{Kingma and Ba}{2014}]%
        {Kingma2014AdamAM}
\bibfield{author}{\bibinfo{person}{Diederik~P. Kingma} {and}
  \bibinfo{person}{Jimmy Ba}.} \bibinfo{year}{2014}\natexlab{}.
\newblock \showarticletitle{Adam: A Method for Stochastic Optimization}.
\newblock \bibinfo{journal}{{\em CoRR\/}}  \bibinfo{volume}{abs/1412.6980}
  (\bibinfo{year}{2014}).
\newblock


\bibitem[\protect\citeauthoryear{Linden, Smith, and York}{Linden
  et~al\mbox{.}}{2003}]%
        {1167344}
\bibfield{author}{\bibinfo{person}{G. Linden}, \bibinfo{person}{B. Smith},
  {and} \bibinfo{person}{J. York}.} \bibinfo{year}{2003}\natexlab{}.
\newblock \showarticletitle{Amazon.com recommendations: item-to-item
  collaborative filtering}.
\newblock \bibinfo{journal}{{\em IEEE Internet Computing\/}}
  \bibinfo{volume}{7}, \bibinfo{number}{1} (\bibinfo{date}{Jan}
  \bibinfo{year}{2003}), \bibinfo{pages}{76--80}.
\newblock
\showISSN{1089-7801}
\showDOI{%
\url{https://doi.org/10.1109/MIC.2003.1167344}}


\bibitem[\protect\citeauthoryear{Oord, Dieleman, and Schrauwen}{Oord
  et~al\mbox{.}}{2013}]%
        {Oord:2013:DCM:2999792.2999907}
\bibfield{author}{\bibinfo{person}{A\"{a}ron van~den Oord},
  \bibinfo{person}{Sander Dieleman}, {and} \bibinfo{person}{Benjamin
  Schrauwen}.} \bibinfo{year}{2013}\natexlab{}.
\newblock \showarticletitle{Deep Content-based Music Recommendation}. In
  \bibinfo{booktitle}{{\em Proceedings of the 26th International Conference on
  Neural Information Processing Systems}} {\em (\bibinfo{series}{NIPS'13})}.
  \bibinfo{publisher}{Curran Associates Inc.}, \bibinfo{address}{USA},
  \bibinfo{pages}{2643--2651}.
\newblock
\showURL{%
\url{http://dl.acm.org/citation.cfm?id=2999792.2999907}}


\bibitem[\protect\citeauthoryear{Salakhutdinov, Mnih, and Hinton}{Salakhutdinov
  et~al\mbox{.}}{2007}]%
        {Salakhutdinov:2007:RBM:1273496.1273596}
\bibfield{author}{\bibinfo{person}{Ruslan Salakhutdinov},
  \bibinfo{person}{Andriy Mnih}, {and} \bibinfo{person}{Geoffrey Hinton}.}
  \bibinfo{year}{2007}\natexlab{}.
\newblock \showarticletitle{Restricted Boltzmann Machines for Collaborative
  Filtering}. In \bibinfo{booktitle}{{\em Proceedings of the 24th International
  Conference on Machine Learning}} {\em (\bibinfo{series}{ICML '07})}.
  \bibinfo{publisher}{ACM}, \bibinfo{address}{New York, NY, USA},
  \bibinfo{pages}{791--798}.
\newblock
\showISBNx{978-1-59593-793-3}
\showDOI{%
\url{https://doi.org/10.1145/1273496.1273596}}


\bibitem[\protect\citeauthoryear{Sarwar, Karypis, Konstan, and Riedl}{Sarwar
  et~al\mbox{.}}{2001}]%
        {Sarwar:2001:ICF:371920.372071}
\bibfield{author}{\bibinfo{person}{Badrul Sarwar}, \bibinfo{person}{George
  Karypis}, \bibinfo{person}{Joseph Konstan}, {and} \bibinfo{person}{John
  Riedl}.} \bibinfo{year}{2001}\natexlab{}.
\newblock \showarticletitle{Item-based Collaborative Filtering Recommendation
  Algorithms}. In \bibinfo{booktitle}{{\em Proceedings of the 10th
  International Conference on World Wide Web}} {\em (\bibinfo{series}{WWW
  '01})}. \bibinfo{publisher}{ACM}, \bibinfo{address}{New York, NY, USA},
  \bibinfo{pages}{285--295}.
\newblock
\showISBNx{1-58113-348-0}
\showDOI{%
\url{https://doi.org/10.1145/371920.372071}}


\bibitem[\protect\citeauthoryear{Shani, Heckerman, and Brafman}{Shani
  et~al\mbox{.}}{2005}]%
        {Shani:2005:MRS:1046920.1088715}
\bibfield{author}{\bibinfo{person}{Guy Shani}, \bibinfo{person}{David
  Heckerman}, {and} \bibinfo{person}{Ronen~I. Brafman}.}
  \bibinfo{year}{2005}\natexlab{}.
\newblock \showarticletitle{An MDP-Based Recommender System}.
\newblock \bibinfo{journal}{{\em J. Mach. Learn. Res.\/}}  \bibinfo{volume}{6}
  (\bibinfo{date}{Dec.} \bibinfo{year}{2005}), \bibinfo{pages}{1265--1295}.
\newblock
\showISSN{1532-4435}
\showURL{%
\url{http://dl.acm.org/citation.cfm?id=1046920.1088715}}


\bibitem[\protect\citeauthoryear{Song, Elkahky, and He}{Song
  et~al\mbox{.}}{2016}]%
        {Song:2016:MDL:2911451.2914726}
\bibfield{author}{\bibinfo{person}{Yang Song}, \bibinfo{person}{Ali~Mamdouh
  Elkahky}, {and} \bibinfo{person}{Xiaodong He}.}
  \bibinfo{year}{2016}\natexlab{}.
\newblock \showarticletitle{Multi-Rate Deep Learning for Temporal
  Recommendation}. In \bibinfo{booktitle}{{\em Proceedings of the 39th
  International ACM SIGIR Conference on Research and Development in Information
  Retrieval}} {\em (\bibinfo{series}{SIGIR '16})}. \bibinfo{publisher}{ACM},
  \bibinfo{address}{New York, NY, USA}, \bibinfo{pages}{909--912}.
\newblock
\showISBNx{978-1-4503-4069-4}
\showDOI{%
\url{https://doi.org/10.1145/2911451.2914726}}


\bibitem[\protect\citeauthoryear{Tan, Xu, and Liu}{Tan et~al\mbox{.}}{2016}]%
        {TanSessRec}
\bibfield{author}{\bibinfo{person}{Yong~Kiam Tan}, \bibinfo{person}{Xinxing
  Xu}, {and} \bibinfo{person}{Yong Liu}.} \bibinfo{year}{2016}\natexlab{}.
\newblock \showarticletitle{Improved Recurrent Neural Networks for
  Session-based Recommendations}. In \bibinfo{booktitle}{{\em Proceedings of
  the 1st Workshop on Deep Learning for Recommender Systems}} {\em
  (\bibinfo{series}{DLRS 2016})}. \bibinfo{publisher}{ACM},
  \bibinfo{address}{New York, NY, USA}, \bibinfo{pages}{17--22}.
\newblock
\showISBNx{978-1-4503-4795-2}
\showDOI{%
\url{https://doi.org/10.1145/2988450.2988452}}


\bibitem[\protect\citeauthoryear{Twardowski}{Twardowski}{2016}]%
        {Twardowski:2016:MCI:2959100.2959162}
\bibfield{author}{\bibinfo{person}{Bartlomiej Twardowski}.}
  \bibinfo{year}{2016}\natexlab{}.
\newblock \showarticletitle{Modelling Contextual Information in Session-Aware
  Recommender Systems with Neural Networks}. In \bibinfo{booktitle}{{\em
  Proceedings of the 10th ACM Conference on Recommender Systems}} {\em
  (\bibinfo{series}{RecSys '16})}. \bibinfo{publisher}{ACM},
  \bibinfo{address}{New York, NY, USA}, \bibinfo{pages}{273--276}.
\newblock
\showISBNx{978-1-4503-4035-9}
\showDOI{%
\url{https://doi.org/10.1145/2959100.2959162}}


\bibitem[\protect\citeauthoryear{Wang, Wang, and Yeung}{Wang
  et~al\mbox{.}}{2015}]%
        {Wang:2015:CDL:2783258.2783273}
\bibfield{author}{\bibinfo{person}{Hao Wang}, \bibinfo{person}{Naiyan Wang},
  {and} \bibinfo{person}{Dit-Yan Yeung}.} \bibinfo{year}{2015}\natexlab{}.
\newblock \showarticletitle{Collaborative Deep Learning for Recommender
  Systems}. In \bibinfo{booktitle}{{\em Proceedings of the 21th ACM SIGKDD
  International Conference on Knowledge Discovery and Data Mining}} {\em
  (\bibinfo{series}{KDD '15})}. \bibinfo{publisher}{ACM}, \bibinfo{address}{New
  York, NY, USA}, \bibinfo{pages}{1235--1244}.
\newblock
\showISBNx{978-1-4503-3664-2}
\showDOI{%
\url{https://doi.org/10.1145/2783258.2783273}}


\bibitem[\protect\citeauthoryear{Yi, Hong, Zhong, Liu, and Rajan}{Yi
  et~al\mbox{.}}{2014}]%
        {YiBeyondClicks}
\bibfield{author}{\bibinfo{person}{Xing Yi}, \bibinfo{person}{Liangjie Hong},
  \bibinfo{person}{Erheng Zhong}, \bibinfo{person}{Nanthan~Nan Liu}, {and}
  \bibinfo{person}{Suju Rajan}.} \bibinfo{year}{2014}\natexlab{}.
\newblock \showarticletitle{Beyond Clicks: Dwell Time for Personalization}. In
  \bibinfo{booktitle}{{\em Proceedings of the 8th ACM Conference on Recommender
  Systems}} {\em (\bibinfo{series}{RecSys '14})}. \bibinfo{publisher}{ACM},
  \bibinfo{address}{New York, NY, USA}, \bibinfo{pages}{113--120}.
\newblock
\showISBNx{978-1-4503-2668-1}
\showDOI{%
\url{https://doi.org/10.1145/2645710.2645724}}


\end{thebibliography}

\end{document}